\documentclass[11pt]{article}
\usepackage[left=3.1cm, right=3.2cm, top=3.5cm, bottom=3.25cm]{geometry}

\usepackage{cite}
\usepackage{amsmath,amssymb,amsfonts}
\usepackage{algorithmic}
\usepackage{graphicx}
\usepackage{textcomp}
\usepackage{xcolor}
\def\BibTeX{{\rm B\kern-.05em{\sc i\kern-.025em b}\kern-.08em
    T\kern-.1667em\lower.7ex\hbox{E}\kern-.125emX}}

\usepackage{tabularx}
\usepackage{tikz,pgfplots}
\usetikzlibrary{calc,patterns,angles,quotes}
\usepackage{mathtools}
\usepackage{xfrac}
\usepackage{siunitx}

\graphicspath{{./figures/}}

\graphicspath{{./figures/}}

\newcounter{problemC}

\usepackage[capitalize,nameinlink,compress]{cleveref}
\crefname{appendix}{Appendix}{Appendices} 
\crefname{figure}{Figure}{Figures} 
\crefname{equation}{}{} 

\crefname{claim}{Claim}{Claims}
\crefname{example}{Example}{Examples}
\crefname{remark}{Remark}{Remarks}
\crefname{assumption}{Assumption}{Assumptions}
\crefname{fact}{Fact}{Facts}
\crefname{proposition}{Proposition}{Propositions}
\crefname{corollary}{Corollary}{Corollaries}
\crefname{lemma}{Lemma}{Lemmas}
\crefname{theorem}{Theorem}{Theorems}
\crefname{definition}{Definition}{Definitions}

\crefformat{problemC}{#2Problem~\ref{prob::#1}#3}
\crefmultiformat{problemC}{Problems~#2\cref{prob::#1}#3}{ and #2\cref{prob::#1}#3}{, #2\cref{prob::#1}#3}{ and #2\cref{prob::#1}#3}

\newcommand{\mc}{\mathcal}

\newcommand{\pcc}{\mathrm{pcc}}
\newcommand{\dc}{\mathrm{dc}}

\newcommand{\ac}{\mathrm{ac}}

\newcommand{\ol}[1]{\overline{#1}}

\newcommand{\B}[1]{\boldsymbol{#1}}
\newcommand{\tran}{\top}

\newcommand{\dq}{\mathrm{dq}}
\newcommand{\dcom}{\mathrm{d}}
\newcommand{\qcom}{\mathrm{q}}

\newcommand{\ab}[1]{\alpha\beta}

\newcommand{\subjto}{\textrm{subject to}}
\allowdisplaybreaks

\DeclareSIUnit \pu {pu}
\DeclareSIUnit \voltampere {VA}

\usepackage[footnotesize]{subfigure}

\title{\Large\bf Frequency Constrained MPC for Efficient Grid Side Operation of Wind Power Conversion Systems
}
\author{{Orcun Karaca}\\
	\textit{ABB Corporate Research Center} \\
	Baden-Dättwil, Switzerland \\
	orcun.karaca@ch.abb.com\\
	\and
	{Georgios Darivianakis}\\
	\textit{Alpiq AG} \\
	Olten, Switzerland \\
	georgios.darivianakis@alpiq.com
}
\date{June 2024}
\begin{document}
	\maketitle

\begin{abstract}
Model predictive control (MPC) has proven its applicability in power conversion control with its fast dynamic response to reference changes while ensuring critical system constraints are satisfied. Even then, the computational burden still remains a challenge for many MPC variants. In this regard, this paper formulates an indirect MPC scheme for grid-side wind converters. A quadratic program with linear constraints is solved in a receding horizon fashion with a subsequent PWM modulator. To facilitate its solution within a few hundreds of microseconds, its decision variables (modulating signals) are restricted to a specific frequency content. This approach limits the increase in problem size due to horizon length. In case studies, the proposed MPC exhibits fast response in faults and operates the converter within its safety limits.\\

\end{abstract}

\section{Introduction}
Multi-phase medium-voltage generators are increasingly employed in offshore wind parks in order to maximize the generated power per wind turbine and thus minimize the cost per MW of installed power~\cite{tsoumas2023elimination}.
Often in such medium-voltage systems, a back-to-back converter configuration is used. One converter system is connected to the multi-phase generator and delivers power to the dc-link(s), whereas a second converter system delivers the power to grid via a multi-phase transformer. Control of these medium-voltage systems is challenging because of the low device switching frequencies. In particular, the design of traditional linear controllers, such as PI controllers, can become complicated due to both the ripple components and the multiple-input multiple-output (MIMO) nature of the multi-phase system.

Given the recent developments in mathematical optimization techniques and their application on embedded systems, MPC has become applicable to fast linear dynamics found in power conversion systems~\cite{karamanakos2020model,harbi2023model,rodriguez2021latest,rodriguez2021latest2,zafra2023long}.
It also further gained popularity with simplicity of its design and its MIMO capability providing flexibility in various operational conditions.  In terms of the decision variable, the MPC methods can be categorized {as either \emph{direct} or \emph{indirect}}. {As the name implies, direct MPC directly manipulates the switching signal of a converter, thus combining control and modulation into a single stage.} Majority of such methods provide high dynamic performance; however, they produce a wide nondeterministic harmonic spectrum, making them unsuitable to grid-connected converter applications that require meeting grid codes, e.g., the IEEE 519 standard.
This can be alleviated using the direct control of optimized pulse patterns with a predetermined harmonic spectrum, as in \cite{dorfling2022generalized,dorfling2022generalized2,begh2022gradient,begh2022gradient2}.

Indirect MPC, on the other hand, {manipulates a reference signal of a modulator that generates the switching signals}~\cite{bolognani2008design,mariethoz2008explicit}. The work in \cite{darivianakis2014model} was the first MPC scheme in the literature combining an upper-layer quadratic program to control currents with a PWM modulator, with an MMC application on hand. This work was later extended to LCL filter dynamics in~\cite{rossi2022indirect}. The use of PWM ensures effectively a constant switching frequency and a {well-defined spectrum with only odd non-triplen harmonics}. Moreover, the upper layer allows for explicit time-domain state constraints in a straightforward manner when compared to the methods available for direct MPC schemes~\cite{keusch2023long}. While it is known that the prediction horizon can bring in performance benefits to the MPC variants in converter control~\cite{karamanakos2014direct}, the work in \cite{darivianakis2014model} had to restrict the prediction horizon to 5--10 steps to keep the problem size (or equivalently, the number of decision variables) moderate.  The main motivation of this paper is to propose an indirect MPC scheme with a drastic decrease in its number of decision variables. \looseness=-1

Our contribution is to formulate an MPC scheme for the grid-side control of a dual-converter wind power conversion system. To enable the fast computation of a solution, we restrict the decision variables to a specific frequency content, which then makes the number of decision variables mostly independent of the length of the prediction horizon. Instead, the number of decision variables will scale linearly with the number of frequency components included. To the best of our knowledge, this is the first investigation of an MPC scheme that achieves a size reduction via frequency component constraints on its decision variables. This approach allows keeping the size of the MPC problem moderate even when long horizons are utilized. 

\looseness=-1

\begin{figure}[t!]
	\centering
	\resizebox{0.9\linewidth}{!}{
		\input{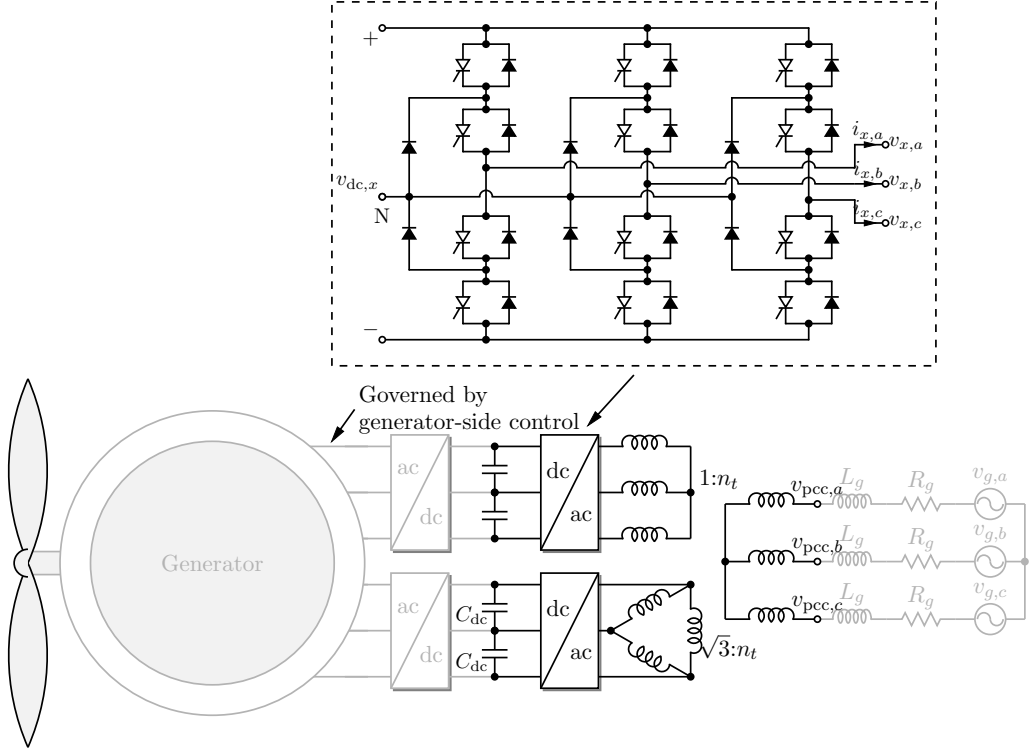}}
	\caption{Wind converter system comprising a dual conversion line connected to the grid through a transformer.}
	\label{fig::system}
\end{figure}

\section{Preliminaries on system modeling}\label{sec::Prel}

Consider a wind power conversion system that comprises a dual conversion line as depicted in the schematic in \cref{fig::system}. On the generator-side, the system is connected to a permanent-magnet synchronous generator with two sets of three-phase windings. On the grid-side, the connection is via a transformer in a Wye-Delta-Wye configuration. Such a configuration has the advantage of eliminating the $ 6n\pm 1\, (n=1, 3, \ldots) $ harmonics when a phase-shift of $ \pi/6\, $rad is applied between the modulating signals of the two lines. 
\looseness=-1

The goal is to develop a control scheme that efficiently operates the grid-side. To this end, we make the following assumptions for system modeling.
The generator-side converters are approximated by controlled current sources that depend on the power produced in the generator.
The control unit of the generator-side converters is, among others, responsible for controlling the neutral-point potential of each conversion line, that is, balancing the upper and lower halves of the dc-link capacitors of each line.
In the following, we present the modeling principles for the grid-side part of the system under consideration, which includes the dc-link capacitors, the inverter units, the transformer and the grid itself. 

The standard three-phase $abc$ frame is indicated as $\B{\xi}_{abc} = [\xi_a~\xi_b~\xi_c]^\tran$. Ac quantities in~$abc$ are always indicated by their subscript. The dynamical equations are based on the $\alpha\beta$-reference frame (also known as the {stationary/orthogonal reference frame}).
The $\alpha\beta$-reference frame is denoted by $\B{\xi}_{\alpha\beta} = [\xi_\alpha~\xi_\beta]^\tran$, and
$\B{\xi}_{\alpha\beta} = \B{K}\B{\xi}_{abc},$
where the {Clarke} matrix is defined as $$\B{K} = \frac{2}{3}\begin{bmatrix}
	1 & -\frac{1}{2} & -\frac{1}{2} \\
	0 & \frac{\sqrt{3}}{2} & -\frac{\sqrt{3}}{2}
\end{bmatrix}.$$
Similarly, the inverse transformation is defined as
$\B{\xi}_{abc}  = \B{K}^{-1}\B{\xi}_{\alpha\beta},$
where the (pseudo) inverse of $\B{K}$ is
$\B{K}^{-1} = \frac{3}{2}\B{K}^{\tran}.$
Whenever the subscript is dropped, $\alpha\beta$-reference frame is to be assumed,  $\B{\xi} = \B{\xi}_{\alpha\beta}$. 
Time dependency of the continuous dynamics and their discretization are dropped for convenience. \looseness=-1

\clearpage

\subsection{Transformer model}

\begin{figure}[t!]
	\centering
	\resizebox{0.75\linewidth}{!}{
		\begin{tikzpicture}[scale=2.54]%
\ifx\dpiclw\undefined\newdimen\dpiclw\fi
\global\def\dpicdraw{\draw[line width=\dpiclw]}
\global\def\dpicstop{;}
\dpiclw=0.8bp
\dpiclw=0.8bp
\dpicdraw[fill=black](0,0) circle (0.007874in)\dpicstop
\draw (0.01,0) node[above=-2bp]{$\B{v}_1$};
\dpicdraw[fill=black](0,-0.5625) circle (0.007874in)\dpicstop
\draw (0.01,-0.5625) node[above=-2bp]{$\B{v}_2$};
\dpicdraw (0,0)
 --(0.25,0)
 --(0.270833,0.041667)
 --(0.3125,-0.041667)
 --(0.354167,0.041667)
 --(0.395833,-0.041667)
 --(0.4375,0.041667)
 --(0.479167,-0.041667)
 --(0.5,0)
 --(0.75,0)\dpicstop
\draw (0.375,0.041667) node[above=-2bp]{$ R_s$};
\dpicdraw (0.75,0)
 --(1,0)\dpicstop
\dpicdraw[line width=0.4bp](1,0) circle (0.00109in)\dpicstop
\dpicdraw (1,0)
 ..controls (1,0.034375) and (1.014625,0.0625)
 ..(1.0325,0.0625)
 ..controls (1.050375,0.0625) and (1.065,0.042813)
 ..(1.065,0.01875)
 ..controls (1.065,-0.005313) and (1.05825,-0.025)
 ..(1.05,-0.025)
 ..controls (1.04175,-0.025) and (1.035,-0.005313)
 ..(1.035,0.01875)
 ..controls (1.035,0.042813) and (1.053,0.0625)
 ..(1.075,0.0625)
 ..controls (1.097,0.0625) and (1.115,0.042813)
 ..(1.115,0.01875)
 ..controls (1.115,-0.005313) and (1.10825,-0.025)
 ..(1.1,-0.025)
 ..controls (1.09175,-0.025) and (1.085,-0.005313)
 ..(1.085,0.01875)
 ..controls (1.085,0.042813) and (1.103,0.0625)
 ..(1.125,0.0625)
 ..controls (1.147,0.0625) and (1.165,0.042813)
 ..(1.165,0.01875)
 ..controls (1.165,-0.005313) and (1.15825,-0.025)
 ..(1.15,-0.025)
 ..controls (1.14175,-0.025) and (1.135,-0.005313)
 ..(1.135,0.01875)
 ..controls (1.135,0.042813) and (1.153,0.0625)
 ..(1.175,0.0625)
 ..controls (1.197,0.0625) and (1.215,0.042813)
 ..(1.215,0.01875)
 ..controls (1.215,-0.005313) and (1.20825,-0.025)
 ..(1.2,-0.025)
 ..controls (1.19175,-0.025) and (1.185,-0.005313)
 ..(1.185,0.01875)
 ..controls (1.185,0.042813) and (1.199625,0.0625)
 ..(1.2175,0.0625)
 ..controls (1.235375,0.0625) and (1.25,0.034375)
 ..(1.25,0)\dpicstop
\dpicdraw[line width=0.4bp](1.25,0) circle (0.00109in)\dpicstop
\dpicdraw (1.25,0)
 --(1.5,0)\dpicstop
\draw (1.125,0.0625) node[above=-2bp]{$ L_s$};
\filldraw (0.86,-0.02125)
 --(0.945,0)
 --(0.86,0.02125) --cycle\dpicstop
\dpicdraw (0.922094,0)
 --(0.86,0)\dpicstop
\draw (0.891047,0) node[above=-2bp]{$ \B{i}_1$};
\dpicdraw[line width=0.4bp](1.5,0) circle (0.00109in)\dpicstop
\dpicdraw (0,-0.5625)
 --(0.25,-0.5625)
 --(0.270833,-0.520833)
 --(0.3125,-0.604167)
 --(0.354167,-0.520833)
 --(0.395833,-0.604167)
 --(0.4375,-0.520833)
 --(0.479167,-0.604167)
 --(0.5,-0.5625)
 --(0.75,-0.5625)\dpicstop
\draw (0.375,-0.520833) node[above=-2bp]{$ R_s$};
\dpicdraw (0.75,-0.5625)
 --(1,-0.5625)\dpicstop
\dpicdraw[line width=0.4bp](1,-0.5625) circle (0.00109in)\dpicstop
\dpicdraw (1,-0.5625)
 ..controls (1,-0.528125) and (1.014625,-0.5)
 ..(1.0325,-0.5)
 ..controls (1.050375,-0.5) and (1.065,-0.519687)
 ..(1.065,-0.54375)
 ..controls (1.065,-0.567813) and (1.05825,-0.5875)
 ..(1.05,-0.5875)
 ..controls (1.04175,-0.5875) and (1.035,-0.567813)
 ..(1.035,-0.54375)
 ..controls (1.035,-0.519687) and (1.053,-0.5)
 ..(1.075,-0.5)
 ..controls (1.097,-0.5) and (1.115,-0.519687)
 ..(1.115,-0.54375)
 ..controls (1.115,-0.567813) and (1.10825,-0.5875)
 ..(1.1,-0.5875)
 ..controls (1.09175,-0.5875) and (1.085,-0.567813)
 ..(1.085,-0.54375)
 ..controls (1.085,-0.519687) and (1.103,-0.5)
 ..(1.125,-0.5)
 ..controls (1.147,-0.5) and (1.165,-0.519687)
 ..(1.165,-0.54375)
 ..controls (1.165,-0.567813) and (1.15825,-0.5875)
 ..(1.15,-0.5875)
 ..controls (1.14175,-0.5875) and (1.135,-0.567813)
 ..(1.135,-0.54375)
 ..controls (1.135,-0.519687) and (1.153,-0.5)
 ..(1.175,-0.5)
 ..controls (1.197,-0.5) and (1.215,-0.519687)
 ..(1.215,-0.54375)
 ..controls (1.215,-0.567813) and (1.20825,-0.5875)
 ..(1.2,-0.5875)
 ..controls (1.19175,-0.5875) and (1.185,-0.567813)
 ..(1.185,-0.54375)
 ..controls (1.185,-0.519687) and (1.199625,-0.5)
 ..(1.2175,-0.5)
 ..controls (1.235375,-0.5) and (1.25,-0.528125)
 ..(1.25,-0.5625)\dpicstop
\dpicdraw[line width=0.4bp](1.25,-0.5625) circle (0.00109in)\dpicstop
\dpicdraw (1.25,-0.5625)
 --(1.5,-0.5625)\dpicstop
\draw (1.125,-0.5) node[above=-2bp]{$ L_s$};
\filldraw (0.86,-0.58375)
 --(0.945,-0.5625)
 --(0.86,-0.54125) --cycle\dpicstop
\dpicdraw (0.922094,-0.5625)
 --(0.86,-0.5625)\dpicstop
\draw (0.891047,-0.5625) node[above=-2bp]{$ \B{i}_2$};
\dpicdraw[line width=0.4bp](1.5,-0.5625) circle (0.00109in)\dpicstop
\dpicdraw (1.5,0)
 --(1.5,-0.5625)\dpicstop
\dpicdraw (1.5,-0.28125)
 --(1.75,-0.28125)
 --(1.770833,-0.239583)
 --(1.8125,-0.322917)
 --(1.854167,-0.239583)
 --(1.895833,-0.322917)
 --(1.9375,-0.239583)
 --(1.979167,-0.322917)
 --(2,-0.28125)
 --(2.25,-0.28125)\dpicstop
\draw (1.875,-0.239583) node[above=-2bp]{$ R_p$};
\filldraw (2.055,-0.3025)
 --(2.14,-0.28125)
 --(2.055,-0.26) --cycle\dpicstop
\dpicdraw (2.117094,-0.28125)
 --(2.055,-0.28125)\dpicstop
\draw (2.086047,-0.28125) node[left=-4bp, above=0bp]{$ \B{i}_\pcc$};
\dpicdraw (2.25,-0.28125)
 --(2.425,-0.28125)\dpicstop
\dpicdraw[line width=0.4bp](2.425,-0.28125) circle (0.00109in)\dpicstop
\dpicdraw (2.425,-0.28125)
 ..controls (2.425,-0.246875) and (2.439625,-0.21875)
 ..(2.4575,-0.21875)
 ..controls (2.475375,-0.21875) and (2.49,-0.238438)
 ..(2.49,-0.2625)
 ..controls (2.49,-0.286563) and (2.48325,-0.30625)
 ..(2.475,-0.30625)
 ..controls (2.46675,-0.30625) and (2.46,-0.286563)
 ..(2.46,-0.2625)
 ..controls (2.46,-0.238438) and (2.478,-0.21875)
 ..(2.5,-0.21875)
 ..controls (2.522,-0.21875) and (2.54,-0.238438)
 ..(2.54,-0.2625)
 ..controls (2.54,-0.286563) and (2.53325,-0.30625)
 ..(2.525,-0.30625)
 ..controls (2.51675,-0.30625) and (2.51,-0.286563)
 ..(2.51,-0.2625)
 ..controls (2.51,-0.238438) and (2.528,-0.21875)
 ..(2.55,-0.21875)
 ..controls (2.572,-0.21875) and (2.59,-0.238438)
 ..(2.59,-0.2625)
 ..controls (2.59,-0.286563) and (2.58325,-0.30625)
 ..(2.575,-0.30625)
 ..controls (2.56675,-0.30625) and (2.56,-0.286563)
 ..(2.56,-0.2625)
 ..controls (2.56,-0.238438) and (2.578,-0.21875)
 ..(2.6,-0.21875)
 ..controls (2.622,-0.21875) and (2.64,-0.238438)
 ..(2.64,-0.2625)
 ..controls (2.64,-0.286563) and (2.63325,-0.30625)
 ..(2.625,-0.30625)
 ..controls (2.61675,-0.30625) and (2.61,-0.286563)
 ..(2.61,-0.2625)
 ..controls (2.61,-0.238438) and (2.624625,-0.21875)
 ..(2.6425,-0.21875)
 ..controls (2.660375,-0.21875) and (2.675,-0.246875)
 ..(2.675,-0.28125)\dpicstop
\dpicdraw[line width=0.4bp](2.675,-0.28125) circle (0.00109in)\dpicstop
\dpicdraw (2.675,-0.28125)
 --(2.85,-0.28125)\dpicstop
\draw (2.55,-0.21875) node[above=-2bp]{$ L_p$};
\definecolor{lcspec}{rgb}{0.70000,0.70000,0.70000}%
\color[rgb]{0.70000,0.70000,0.70000}%
\global\let\dpiclidraw=\dpicdraw\global\let\dpicfidraw=\filldraw
\global\def\dpicdraw{\dpiclidraw[color=lcspec]}
\global\def\filldraw{\dpicfidraw[color=lcspec]}
\dpicdraw (2.85,-0.28125)
 --(3.025,-0.28125)
 --(3.045833,-0.239583)
 --(3.0875,-0.322917)
 --(3.129167,-0.239583)
 --(3.170833,-0.322917)
 --(3.2125,-0.239583)
 --(3.254167,-0.322917)
 --(3.275,-0.28125)
 --(3.45,-0.28125)\dpicstop
\draw (3.15,-0.239583) node[above=-2bp]{$ R_g$};
\dpicdraw (3.45,-0.28125)
 --(3.625,-0.28125)\dpicstop
\dpicdraw[line width=0.4bp](3.625,-0.28125) circle (0.00109in)\dpicstop
\dpicdraw (3.625,-0.28125)
 ..controls (3.625,-0.246875) and (3.639625,-0.21875)
 ..(3.6575,-0.21875)
 ..controls (3.675375,-0.21875) and (3.69,-0.238438)
 ..(3.69,-0.2625)
 ..controls (3.69,-0.286563) and (3.68325,-0.30625)
 ..(3.675,-0.30625)
 ..controls (3.66675,-0.30625) and (3.66,-0.286563)
 ..(3.66,-0.2625)
 ..controls (3.66,-0.238438) and (3.678,-0.21875)
 ..(3.7,-0.21875)
 ..controls (3.722,-0.21875) and (3.74,-0.238438)
 ..(3.74,-0.2625)
 ..controls (3.74,-0.286563) and (3.73325,-0.30625)
 ..(3.725,-0.30625)
 ..controls (3.71675,-0.30625) and (3.71,-0.286563)
 ..(3.71,-0.2625)
 ..controls (3.71,-0.238438) and (3.728,-0.21875)
 ..(3.75,-0.21875)
 ..controls (3.772,-0.21875) and (3.79,-0.238438)
 ..(3.79,-0.2625)
 ..controls (3.79,-0.286563) and (3.78325,-0.30625)
 ..(3.775,-0.30625)
 ..controls (3.76675,-0.30625) and (3.76,-0.286563)
 ..(3.76,-0.2625)
 ..controls (3.76,-0.238438) and (3.778,-0.21875)
 ..(3.8,-0.21875)
 ..controls (3.822,-0.21875) and (3.84,-0.238438)
 ..(3.84,-0.2625)
 ..controls (3.84,-0.286563) and (3.83325,-0.30625)
 ..(3.825,-0.30625)
 ..controls (3.81675,-0.30625) and (3.81,-0.286563)
 ..(3.81,-0.2625)
 ..controls (3.81,-0.238438) and (3.824625,-0.21875)
 ..(3.8425,-0.21875)
 ..controls (3.860375,-0.21875) and (3.875,-0.246875)
 ..(3.875,-0.28125)\dpicstop
\dpicdraw[line width=0.4bp](3.875,-0.28125) circle (0.00109in)\dpicstop
\dpicdraw (3.875,-0.28125)
 --(4.05,-0.28125)\dpicstop
\draw (3.75,-0.21875) node[above=-2bp]{$ L_g$};
\definecolor{fcspec}{rgb}{0.70000,0.70000,0.70000}%
\global\def\dpicstop{--}
\global\let\dpicfdraw=\dpicdraw\global\def\dpicdraw{}
\path[fill=fcspec]
\dpicdraw (4.05,-0.28125) circle (0.007874in)\dpicstop
cycle; \global\let\dpicdraw=\dpicfdraw\global\def\dpicstop{;}
\dpicdraw (4.05,-0.28125) circle (0.007874in)\dpicstop
\draw (4.06,-0.28125) node[above=-2bp]{$\B{v}_{g}$};
\color{black}\global\let\dpicdraw=\dpiclidraw%
\global\let\filldraw=\dpicfidraw
\dpicdraw[fill=black](2.85,-0.28125) circle (0.007874in)\dpicstop
\draw (2.86,-0.28125) node[above=-2bp]{$\B{v}_{\pcc}$};
\dpicdraw[fill=black](1.5,-0.28125) circle (0.007874in)\dpicstop
\end{tikzpicture}%
}
	\caption{Transformer equivalent (per-unit) circuit of a dual converter system in the $\alpha\beta$-reference frame.}
	\label{fig::equivalent_circuit}
\end{figure}
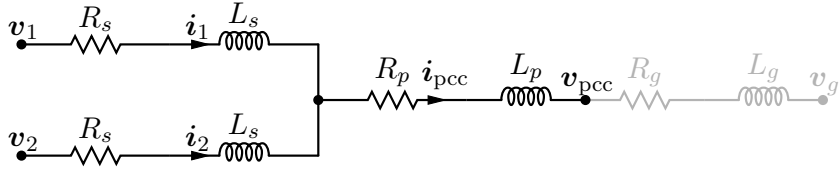

A converter is denoted with $x \in \{1,2\}$. Define $\B{v}_{x,abc} = [v_{x,a}~v_{x,b}~v_{x,c}]^\tran$ as the three-phase voltages of converter $x$.
The output voltages of the converters, denoted by $\B{v}_x = [v_{x,\alpha}~v_{x,\beta}]^\tran$, are described as
$\B{v}_1 = \B{K}\B{v}_{1,abc},
\B{v}_2 = \B{D}\B{K}\B{v}_{2,abc}$,
where $\B{D} = \B{R}\left(\tfrac{\pi}{6}\right)$
applies {a \ang{30}} rotation due to the Delta configuration of the second conversion line.
\looseness=-1

{The transformer can be described using the equivalent circuit in \cref{fig::equivalent_circuit}.} The differential equations are given by
{\medmuskip=.75mu \thickmuskip=1mu \thinmuskip=1mu\begin{equation}\label{eq::dynamics_v0}
	\begin{array}{l}
		\B{v}_1 - R_s \B{i}_1 - L_s \dfrac{d \B{i}_1}{dt} - R_p (\B{i}_1 + \B{i}_2) - L_p \dfrac{d (\B{i}_1 + \B{i}_2)}{dt} = \B{v}_\pcc, \\[2ex]
		\B{v}_2 - R_s \B{i}_2 - L_s \dfrac{d \B{i}_2}{dt} - R_p (\B{i}_1 + \B{i}_2) - L_p \dfrac{d (\B{i}_1 + \B{i}_2)}{dt} = \B{v}_\pcc,
	\end{array}
\end{equation}}where $\B{v}_\pcc $ is the voltage measured at the point of common coupling (PCC). Grid parameters and grid voltage source will not be estimated, and assumed to be unknown quantities. Potential steady-state offsets due to the unknown model components {and/or disturbances} will later be compensated for by integral action, presented with the overall control scheme.

We work with the average and difference quantities for the dual conversion line system~\cite{karttunen2013decoupled}. This will allow for a convenient way to formulate our objectives.  Thus, define
\begin{equation*}
	\begin{split}
	&\B{v}_\mathrm{av} = \dfrac{(\B{v}_1 + \B{v}_2)}{2},\quad~ \B{i}_\mathrm{av} = \dfrac{(\B{i}_1 + \B{i}_2)}{2},~\\
	&\B{v}_\mathrm{df} = \dfrac{(\B{v}_1 - \B{v}_2)}{2},\quad~ \B{i}_\mathrm{df} = \dfrac{(\B{i}_1 - \B{i}_2)}{2},~\\
	&R_\mathrm{av} = 2 R_p + R_s,\quad~ L_\mathrm{av} = 2 L_p + L_s.\\
		\end{split}
\end{equation*}
{With these definitions}, \cref{eq::dynamics_v0} becomes
\begin{equation}\label{eq::dynamics_v1}
	\begin{split}
	\B{v}_{\mathrm{av}} - R_\mathrm{av} \B{i}_\mathrm{av} - L_\mathrm{av} \dfrac{d \B{i}_\mathrm{av}}{dt} &= \B{v}_\pcc,\\  \B{v}_{\mathrm{df}} - R_s \B{i}_\mathrm{df} - L_s \dfrac{d \B{i}_\mathrm{df}}{dt} &= 0.
		\end{split}
\end{equation}
Observe that the grid-injected current is $\B{i}_\pcc =2\B{i}_\mathrm{av} $.

\subsection{Dc-link dynamics}
Assuming it is purely capacitive,  the energy balance is
\begin{equation}\label{eq::energy_balance_v0}
	\begin{split}
	P_{\dc, 1} - \B{v}_1^\top \B{i}_1 &= \dfrac{1}{2} \dfrac{C_{\dc}}{2} \dfrac{d v_{\dc, 1}^2}{dt},\\
	P_{\dc, 2} - \B{v}_2^\top \B{i}_2 &= \dfrac{1}{2} \dfrac{C_{\dc}}{2} \dfrac{d v_{\dc, 2}^2}{dt},
\end{split}
\end{equation}
where $P_{\dc, x}$ is delivered by the corresponding generator-side converter.
From \cref{eq::energy_balance_v0}, we have that
{\medmuskip=.75mu \thickmuskip=1mu \thinmuskip=1mu\begin{equation}\label{eq::energy_balance_v1}
	\begin{array}{l}
		P_{\dc, 1} + P_{\dc, 2} - (\B{v}_1^\top \B{i}_1 + \B{v}_2^\top \B{i}_2) = \dfrac{1}{2} \dfrac{C_{\dc}}{2} \dfrac{d (v_{\dc, 1}^2 + v_{\dc, 2}^2)}{dt}, \\[2ex]
		P_{\dc, 1} - P_{\dc, 2} - (\B{v}_1^\top \B{i}_1 - \B{v}_2^\top \B{i}_2) = \dfrac{1}{2} \dfrac{C_{\dc}}{2} \dfrac{d (v_{\dc, 1}^2 - v_{\dc, 2}^2)}{dt}.
	\end{array}
\end{equation}}The terms $\B{v}_1^\top \B{i}_1 $ and $ \B{v}_2^\top \B{i}_2$ make \cref{eq::energy_balance_v1} bilinear. Observe that the states, in this case $\B{i}_x$,  are being multiplied with inputs, $\B{v}_x$. {A linear approximation of these bilinear terms result in~\cref{eq::energy_balance_v1} becoming}\looseness=-1
{\medmuskip=.75mu \thickmuskip=1mu \begin{equation*}
	\begin{array}{l}
		P_{\dc, 1} + P_{\dc, 2} - \B{v}_\pcc^\top (\B{i}_1 + \B{i}_2) = \dfrac{1}{2} \dfrac{C_{\dc}}{2} \dfrac{d (v_{\dc, 1}^2 + v_{\dc, 2}^2)}{dt}, \\[2ex]
		P_{\dc, 1} - P_{\dc, 2} - \B{v}_\pcc^\top (\B{i}_1 - \B{i}_2) = \dfrac{1}{2} \dfrac{C_{\dc}}{2} \dfrac{d (v_{\dc, 1}^2 - v_{\dc, 2}^2)}{dt},
	\end{array}
\end{equation*}}or equivalently, when working with the average and difference quantities, 
\begin{equation}\label{eq:bilinear}
	\begin{split}
		&P_{\dc, \mathrm{av}} - \B{v}_\pcc^\top \B{i}_\mathrm{av} = \dfrac{1}{2} \dfrac{C_{\dc}}{2} \dfrac{d v_{\dc, \mathrm{sav}}}{dt},\\
		&P_{\dc, \mathrm{df}} - \B{v}_\pcc^\top \B{i}_\mathrm{df} = \dfrac{1}{2} \dfrac{C_{\dc}}{2} \dfrac{d v_{\dc, \mathrm{sdf}}}{dt},
	\end{split}
\end{equation} with the definitions
\begin{equation*}
	\begin{split}
		&P_{\dc, \mathrm{av}} = \dfrac{P_{\dc, 1} + P_{\dc, 2}}{2},~ P_{\dc, \mathrm{df}} = \dfrac{P_{\dc, 1} - P_{\dc, 2}}{2},\\
		&v_{\dc, \mathrm{sav}} = \dfrac{v_{\dc, 1}^2 + v_{\dc, 2}^2}{2},~ v_{\dc, \mathrm{sdf}} = \dfrac{v_{\dc, 1}^2 - v_{\dc, 2}^2}{2}.
	\end{split}
\end{equation*}
Working with the average and difference quantities for the dc-link dynamics has an advantage: It permits prioritizing control of the average {quantities (which is what the grid operator cares about)} over the difference {quantities (internal converter quantities that are not important to the grid operator)}

\subsection{Grid modeling}
The evolution of $\B{v}_\pcc$ in the presence of unbalanced grid and harmonic distortions will be predicted by considering both the positive and the negative symmetric sequences for each harmonic component. For an overview of symmetrical components, we kindly refer to \cite[Apps A.2 and A.3]{teodorescu_grid_2011}.  In particular, let
$$\B{v}_\pcc =  \sum_{h \in \mc H} \B{v}_{\pcc, h} = \sum_{h \in \mc H} \big(\B{v}_{\pcc, h, ps} + \B{v}_{\pcc, h, ns}\big),$$ where $ \B{v}_{\pcc, h, ps}$ is the positive sequence and $ \B{v}_{\pcc, h, ns}$ is the negative sequence of harmonic $ h $, with $ \mc H $ the set of harmonics under consideration. A phase-locked loop (PLL) together with notch filters or a linear observer can be used to identify these components, see~\cite{kaura1997operation,ho2013method}. For each harmonic $ h \in \mc H $, we have\looseness=-1
\begin{equation}\label{eq::oscillator}
	\begin{array}{l}
	\dfrac{d \B{v}_{\pcc, h, ps}}{dt} = h \omega \B{J} \B{v}_{\pcc, h, ps},\\
	\dfrac{d \B{v}_{\pcc, h, ns}}{dt} = -h \omega \B{J} \B{v}_{\pcc, h, ns},
\end{array}
\end{equation}
where $ \omega $ represents the fundamental frequency of the grid voltage which is identified via a PLL at the PCC, and $\B{J} = \begin{bmatrix}
	0 & -1 \\
	1 & 0
\end{bmatrix}$. The sum of the two components from \cref{eq::oscillator} represents an unbalanced harmonic oscillator at frequency $ h \omega $.

\section{Optimization problem formulation}\label{sec::MPC}

This section formulates the constrained optimization problem \eqref{prob::MPC} to be solved at each iteration of the proposed closed-loop MPC scheme. 

Control objectives can be listed as follows: 
The grid-side converters have to regulate the active power delivery to maintain the dc-link voltage at its reference value. The grid-side converters have to regulate also the difference of the two dc-link voltages of the two conversion lines by controlling the difference current.
The reference for the reactive power is provided to the grid-side converters by the wind-park operator via the positive sequence $\qcom$-current reference (in the {synchronous $\dq$-reference frame}~\cite[\S 8]{teodorescu_grid_2011}). During low-voltage fault-ride-through, this reference is directly computed by a grid-code-defined formula so as to counteract the PCC voltage collapse. 
During asymmetric fault-ride-through, the references for the negative sequence currents are again provided by a grid-code formula in a similar fashion, and then converted to negative sequence active and reactive power references. This power reference tracking approach is particularly helpful since tracking scalar quantities (e.g., active and reactive positive and negative sequence powers) simplifies considerably the MPC problem and allows for integral action later. 
\looseness=-1

In the context of achieving the aforementioned objectives, the frequency-constrained MPC is given by the quadratic program with linear constraints described in \eqref{prob::MPC}:

\begin{equation}\tag{$\mathcal{P}_{\mathrm{MPC}}$}\label{prob::MPC}
	\begin{array}{r@{~~~}l}
		\textrm{minimize} & \sum_{t \in \mathcal T} \Big[ \lambda_{\dc, \mathrm{sav}} (v_{\dc, \mathrm{sav}} - V_{\dc, \mathrm{sav}}^*)^2 + \lambda_{\dc, \mathrm{sdf}} v_{\dc, \mathrm{sdf}}^2 + \lambda_{Q, ps} (2 \B{v}_{\pcc, ps}^\top \B{J} \B{i}_{\mathrm{av}, ps} - Q_{ps}^*)^2\\
		& \hspace{3cm} + \lambda_{P, ns} (2 \B{v}_{\pcc, ns}^\top \B{i}_{\mathrm{av}, ns} - P_{ns}^*)^2 + \lambda_{Q, ns} (2 \B{v}_{\pcc, ns}^\top \B{J} \B{i}_{\mathrm{av}, ns} - Q_{ns}^*)^2 \Big] \vspace{.2cm}\\[2ex]
		\subjto & \B{v}_{\mathrm{av}} =  \B{v}_{\mathrm{av}, ps} + \B{v}_{\mathrm{av}, ns},~~ \frac{d \B{v}_{\mathrm{av}, ps}}{dt} = \omega \B{J} \B{v}_{\mathrm{av}, ps},~~ \frac{d \B{v}_{\mathrm{av}, ns}}{dt} = -\omega \B{J} \B{v}_{\mathrm{av}, ns}, \\[1ex]
		& \B{v}_{\mathrm{df}} =  \B{v}_{\mathrm{df}, ps} ,~~ \frac{d \B{v}_{\mathrm{df}, ps}}{dt} = \omega J \B{v}_{\mathrm{df}, ps}, \\[1ex]
		& \B{v}_\pcc =  \B{v}_{\pcc, ps} + \B{v}_{\pcc, ns},~~ \frac{d \B{v}_{\pcc, ps}}{dt} = \omega \B{J} \B{v}_{\pcc, ps},~~ \frac{d \B{v}_{\pcc, ns}}{dt} = -\omega \B{J} \B{v}_{\pcc, ns},\\[1ex]
		& \B{i}_{\mathrm{av}} =  \B{i}_{\mathrm{av}, ps} + \B{i}_{\mathrm{av}, ns},~~ \frac{d \B{i}_{\mathrm{av}, ps}}{dt} = \omega J \B{i}_{\mathrm{av}, ps},~~ \frac{d \B{i}_{\mathrm{av}, ns}}{dt} = -\omega J i_{\mathrm{av}, ns},\\[1ex]
		& \B{v}_{\mathrm{av}} - R_\mathrm{av} \B{i}_\mathrm{av} - L_\mathrm{av} \frac{d \B{i}_\mathrm{av}}{dt} = \B{v}_\pcc,\\[1ex]
		& \B{v}_{\mathrm{df}} - R_s \B{i}_\mathrm{df} - L_s \frac{d \B{i}_\mathrm{df}}{dt} = 0, \\[1ex]
		& P_{\dc, \mathrm{av}} - \B{v}_\pcc^\top \B{i}_\mathrm{av} = \frac{1}{2} \frac{C_{\dc}}{2} \frac{d v_{\dc, \mathrm{sav}}}{dt},\\ [1ex]
		& P_{\dc, \mathrm{df}} - \B{v}_\pcc^\top \B{i}_\mathrm{df} = \frac{1}{2} \frac{C_{\dc}}{2} \frac{d v_{\dc, \mathrm{sdf}}}{dt}, \\[1ex]
		& \|\B{i}_\mathrm{av} + \B{i}_\mathrm{df}\|_\infty \le \ol{i_{cl}},~~ \|\B{i}_\mathrm{av} + \B{i}_\mathrm{df}\|_1 \le \frac{2}{\sqrt{2}}\ol{i_{cl}},\\[1ex]
		& \|\B{i}_\mathrm{av} - \B{i}_\mathrm{df}\|_\infty \le \ol{i_{cl}},~~ \|\B{i}_\mathrm{av} - \B{i}_\mathrm{df}\|_1 \le \frac{2}{\sqrt{2}}\ol{i_{cl}}, \\[1ex]
		& \|\B{v}_\mathrm{av} + \B{v}_\mathrm{df}\|_\infty \le \ol{v_{cl}},~~\|\B{v}_\mathrm{av} + \B{v}_\mathrm{df}\|_1 \le \frac{2}{\sqrt{2}}\ol{v_{cl}}, \\[1ex]
		& \|\B{v}_\mathrm{av} - \B{v}_\mathrm{df}\|_\infty \le \ol{v_{cl}},~~ \|\B{v}_\mathrm{av} - \B{v}_\mathrm{df}\|_1 \le \frac{2}{\sqrt{2}}\ol{v_{cl}}.  
	\end{array}\vspace{.5cm}
\end{equation}

We use the subscript $ ps $ and $ ns $ to refer to positive and negative sequence fundamental frequency quantities, respectively. For this particular formulation, we do not utilize the higher order harmonics. We used the formulas for active and reactive power: $P = \B{v}^\top \B{i} \text{ and } Q = \B{v}^\top \B{J} \B{i}$. The weights $ \lambda_{\dc, \mathrm{sav}},\, \lambda_{\dc, \mathrm{sdf}},\, \lambda_{Q, ps},$ $\lambda_{P, ns},$ $\lambda_{Q, ns} $ are predefined based on the priority of the different objectives. Moreover, $ \ol{{i}_{cl}} $ and $ \ol{v_{cl}} $ denote the conversion line current and voltage limits.\footnote{Defining these constraints would require using the second-norm (e.g., $ \|\B{i}_\mathrm{av} + \B{i}_\mathrm{df}\|_2 \le \ol{{i}_{cl}} $), which results in a second-order cone. However, this poses a great challenge in short time intervals. Thus, in \eqref{prob::MPC} we use a combination of the infinity- and one-norm to outer approximate the second-norm. In addition, one could include an instantaneous tangent or other precomputed linear constraints to better approximate the second norm. For the voltage constraint, alternatively, a hexagonal constraint could be considered for operation in overmodulation.}

The frequency-constrained MPC receives the following inputs, and they are not decision variables in~\eqref{prob::MPC}: 
\textit{(i)} the reference signal $ V_{\dc, \mathrm{sav}}^* $ for dc-link capacitors voltages;
\textit{(ii)} the reference signal $ Q_{ps}^* $ for positive sequence reactive power injection;
\textit{(iii)} the reference signals $ P_{ns}^* $ and $ Q_{ns}^* $ for negative sequence active and reactive power;
\textit{(iv)} the limits on the line currents, $ \ol{i_{cl}} $, and the voltages $ \ol{v_{cl}} $;
\textit{(v)} the average and difference powers, $ P_{\dc, \mathrm{av}} $ and $ P_{\dc, \mathrm{df}} $, originating from the generator side;
\textit{(vi)} the positive $ \B{v}_{\pcc, ps} $ and negative sequence $ \B{v}_{\pcc, ns} $ of the PCC voltage at time $ t = 0 $;
\textit{(vii)} the average, $ \B{i}_\mathrm{av} $, and difference, $ \B{i}_\mathrm{df} $, currents flowing in the conversion lines at $ t = 0 $. 
As in the indirect MPC schemes, the only decision variables are the manipulated quantities for control; in this particular case, the converter voltages $\B{v}_\mathrm{av}$ and $\B{v}_\mathrm{df}$. Their first time instances are converted into the modulating signals $\B{u}_1^*$ and $\B{u}_2^*$.\looseness=-1

\textbf{Main features:} The decision variables $ \B{v}_\mathrm{av} $ and $ \B{v}_\mathrm{df} $ are frequency constrained. In the case of~\eqref{prob::MPC},  $\B{v}_\mathrm{av}$ possesses positive and negative sequence fundamental components, whereas $\B{v}_\mathrm{df}$ possesses only a positive sequence component. By enforcing the variables $ \B{v}_\mathrm{av} $ and $ \B{v}_\mathrm{df} $ to a specific frequency content, we essentially reduce the MPC decision variables from $2\times2\times|\mathcal{T}|$ to $3\times2$. Notice that once the first time step instances of $\B{v}_{\mathrm{av}, ps}$, $\B{v}_{\mathrm{av}, ns}$, and $\B{v}_{\mathrm{df}, ps}$ are decided upon, all the other quantities for the rest of the prediction steps (e.g., $v_{\dc, \mathrm{sav}}$, $v_{\dc, \mathrm{sdf}}$, $\B{i}_{\mathrm{av},ps}$, $\B{i}_{\mathrm{av},ns}$, $\B{i}_{\mathrm{df}}$ and others) are fully determined. Hence, the number of decision variables is independent of the prediction horizon length.\footnote{In practice, this may not be completely true due to slack variables which are needed to ensure feasibility under all operating conditions. This applies when implementing any MPC scheme. Increasing the prediction horizon would eventually increase the number of slack variables. Note that the weights for the slack variables of the current and voltage limits are chosen to be high to ensure these constraints are not violated even during faults.}  Instead, the problem size scales only linearly with the number of frequency components included. 

To the best of our knowledge, this is the first investigation of an MPC scheme that achieves a size-reduction via frequency component constraints on its decision variables. In addition, measuring $ \B{i}_\mathrm{av} $ at time $ t=0 $ while obtaining the positive $ \B{i}_{\mathrm{av}, ps} $ and negative $ \B{i}_{\mathrm{av}, ns} $ sequence currents within the problem formulation means that the MPC is implicitly using an internal estimation of these quantities based on the system dynamical equations and the provided estimation of the positive and negative sequence PCC voltages. This is an important feature of the proposed scheme since it does not require relying on external resources to estimate the positive and negative sequence of the currents, making it fast in assessing them.\looseness=-1

\section{Overall control scheme}\label{sec::CtrA}

\begin{figure}[t!]
	\centering
	\includegraphics[width=1.075\textwidth]{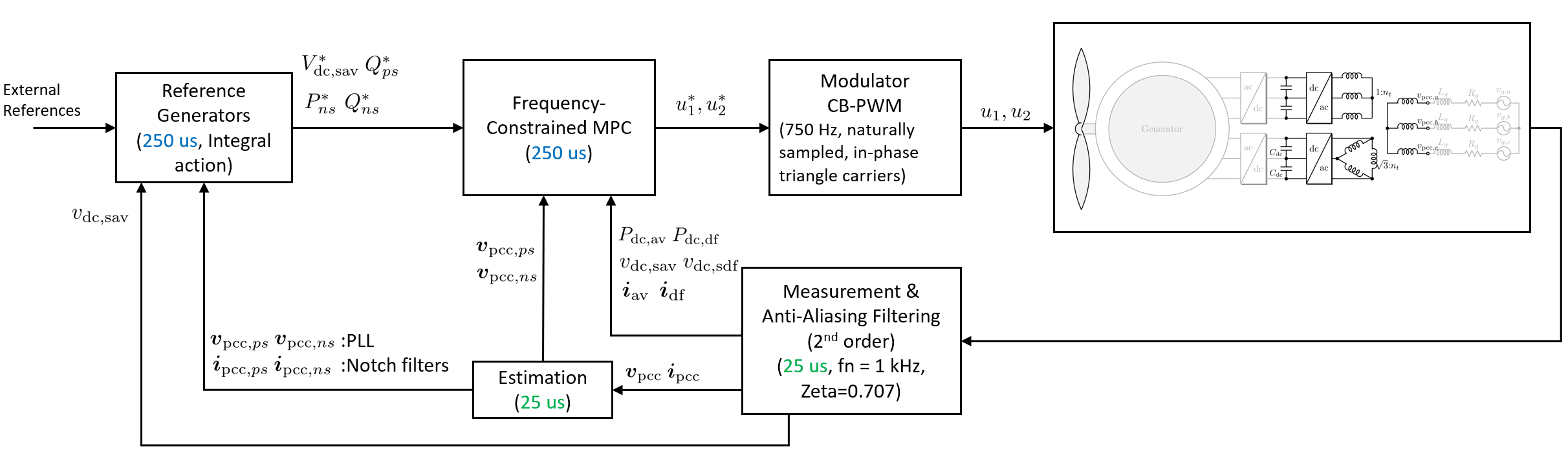}
	\caption{Control structure of frequency constrained MPC.}
	\label{fig::constrolstructure}
\end{figure}

This section discusses the overall control structure. The main blocks of \cref{fig::constrolstructure} are summarized below:

\textbf{Measuring}: We measure the voltage and currents at the PCC, the dc-link and the conversion lines. This happens on fixed $ 25 \,\mu s$ intervals in series with a zero-order-hold (ZOH).
The anti-aliasing filter is also used to filter frequencies above the Nyquist frequency. We account for the delay introduced by the anti-aliasing filter. 

\textbf{Estimation}: The positive and negative sequence PCC voltages are estimated via a PLL-based structure. Moreover, notch-filters are used to obtain the positive and negative sequence grid currents. Note that the positive and negative sequence currents are not directly used in the MPC formulation. They enter the reference generator block for integral action.

\textbf{Reference Generators}: The reference generator block serves a dual purpose:

\textit{(i)} Given the positive sequence $\qcom$-current reference, $ \B{i}_{\pcc, \qcom, ps}^* $, we compute the reactive power reference of the positive sequence $ Q_{\pcc, ps}^* $ as
$$Q_{\pcc, ps}^* = \B{v}_{\pcc, ps}^\top \B{J} \B{R}_{\alpha\beta, \dq}(\omega t) [0, \B{i}_{\pcc, \qcom, ps}^*]^\top,$$
where $ \B{v}_{\pcc, ps} $ is the estimate of the positive sequence voltage at PCC, and $ \B{R}_{\alpha\beta, \dq}(\omega t) $ is the transformation matrix from the rotating $\dq$-reference frame to the stationary $\alpha\beta $. 
Similarly, the references $ \B{i}_{\pcc, \dcom, ns}^*,\, \B{i}_{\pcc, \qcom, ns}^* $ on the negative sequence $\dq$ currents are translated into power references as
$$P_{\pcc, ns}^* = \B{v}_{\pcc, ns}^\top \B{R}_{\alpha\beta, \dq}(\omega t) [\B{i}_{\pcc, \dcom, ns}^*, \B{i}_{\pcc, \qcom, ns}^*]^\top,$$ $$Q_{\pcc, ns}^* = \B{v}_{\pcc, ns}^\top \B{J} \B{R}_{\alpha\beta, \dq}(\omega t) [\B{i}_{\pcc, \dcom, ns}^*, \B{i}_{\pcc, \qcom, ns}^*]^\top.$$

\begin{figure}[t!]
	\centering
	\includegraphics[width=0.55\textwidth]{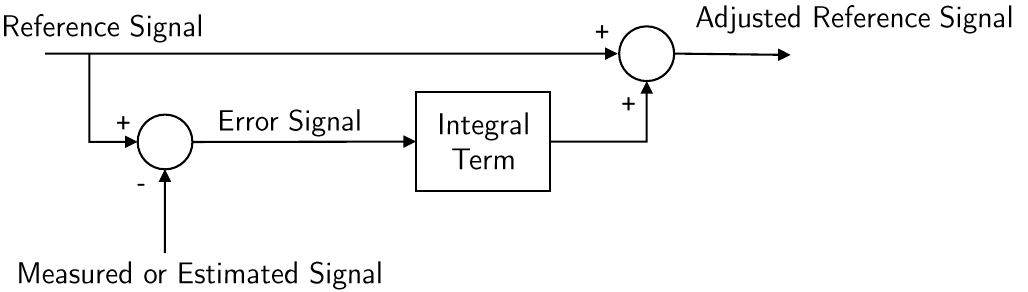}
	\caption{Integral action on MPC scheme by adjusting the reference signal.}
	\label{fig::mpc_integral}
\end{figure}

\textit{(ii)} We provide integral action to compensate for steady-state offsets by adjusting the references as in \cref{fig::mpc_integral}.
Steady state errors can occur due to uncompensated delays, due to model approximations, e.g., as in \cref{eq:bilinear}, or due to unmodeled dynamics such as the grid impedance. Integral action is applied to all the reference signals of the MPC scheme, including the capacitor voltage balancing. To provide integral action to the positive and negative sequence active and reactive power references, the estimates of the positive and negative sequence currents at PCC are used.

\textbf{Frequency Constrained MPC}: This blocks implements MPC in a receding horizon with $ 250 \, \mu s $ discretization and $0.01\, s$ horizon. The internal dynamical model is used to compensate the delays introduced by the measuring, anti-aliasing filter and estimation blocks. In addition to these delays, the delay of executing the MPC block every $ 250 \, \mu s $ is also accounted for. The optimization problem in \cref{prob::MPC} is solved and the first inputs, i.e., the modulating signals of each conversion line, are passed to the respective modulator. \looseness=-1

\begin{figure}[t!]
	\centering
	\includegraphics[width=0.55\textwidth]{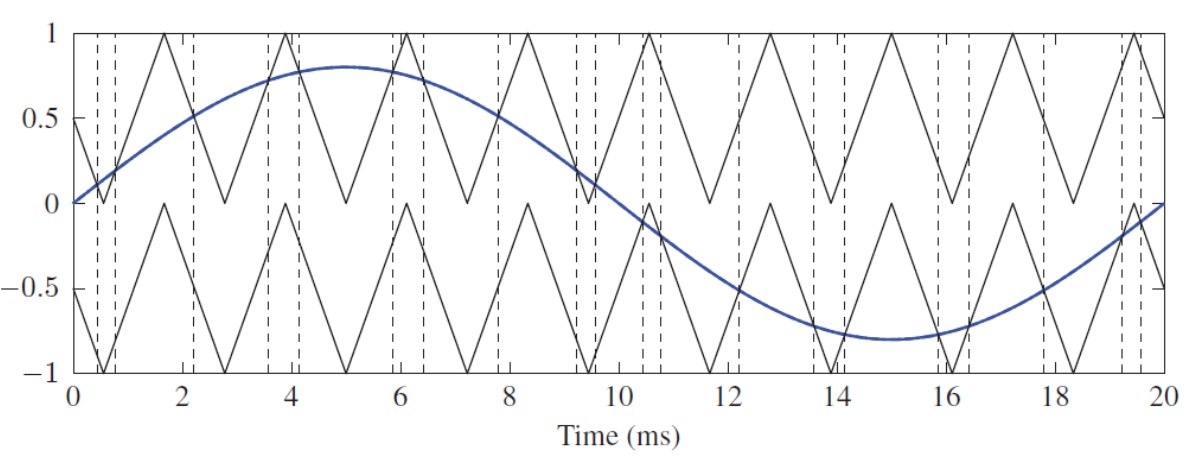}
	\caption{Modulating signal, and upper and lower triangular carrier signals.}
	\label{fig::cbpwm}
\end{figure}

\textbf{Modulator}: A carrier-based pulse width modulator (CB-PWM) is used which is naturally sampled with its triangle carriers in phase. An example with the carrier frequency $ 450\, $Hz is depicted in \cref{fig::cbpwm}. For the case studies, the carrier frequency is chosen as $ 750\, $Hz, resulting in device switching frequencies slightly above $ 350\, $Hz.

\section{Case studies}
In this section, we perform a simulation study to assess the performance of the proposed frequency-constrained MPC scheme. The wind converter system is implemented in Matlab/Simulink, and YALMIP\cite{lofberg2004yalmip} and OSQP\cite{banjac2017embedded,stellato2020osqp} are used to formulate and solve the resulting quadratic program. The rated values of the system are shown in Table~\ref{tab:rated values}. A per-unit system is established for the single converter values. System parameters (e.g., for the transformer and the dc-link capacitor) are confidential. The relevant transformer parameters are $ R_\mathrm{av}=\SI{3.1}{\ohm}$ and $ L_\mathrm{av}=\SI{178}{\milli\henry}$.

We provide two asymmetric faults at short circuit ratio (SCR) 3 happening in the container and the source side to cover a spectrum of fault situations. The SCR 3 case is chosen in particular to show how the integral action can compensate for the modeling mismatch. The proposed method showed a good performance in a wide range of studies (e.g., symmetric faults and/or different grid strengths and XR ratios). Due to space limitations of the conference, these additional studies will not be included in this paper, but they are left out for a future work.

	\begin{table}[t!]
	\caption{Rated values of the medium-voltage converter system.}
	\label{tab:rated values}
	\centering
	\begin{tabular}{lcl}
		\hline
		Parameter & Symbol & Value \\
		\hline
		Rated apparent power of a converter & $S_{\mathrm{R},\mathrm{conv}}$ & $\SI{6}{{\mega\voltampere}}$ \\ 
		Rated apparent power of the transformer & $S_{\mathrm{R},\mathrm{trafo}}$ & $\SI{14}{{\mega\voltampere}}$ \\ 
		Rated voltage at the primary (grid) & $V_{\mathrm{R},\mathrm{prim}}$ & $\SI{66}{\kilo\volt}$ \\
		Rated voltage at the secondary (converter) & $V_{\mathrm{R},\mathrm{sec}}$ & $\SI{3.1}{\kilo\volt}$ \\
		Rated angular frequency &  $\omega_\mathrm{R}$ & $\SI[parse-numbers=false]{2 \pi 50}{\radian\per\second}$ \\
		\hline
	\end{tabular}

\end{table}

\begin{figure}[t!]
	\centering
	\subfigure[]{\includegraphics[width=0.47\textwidth]{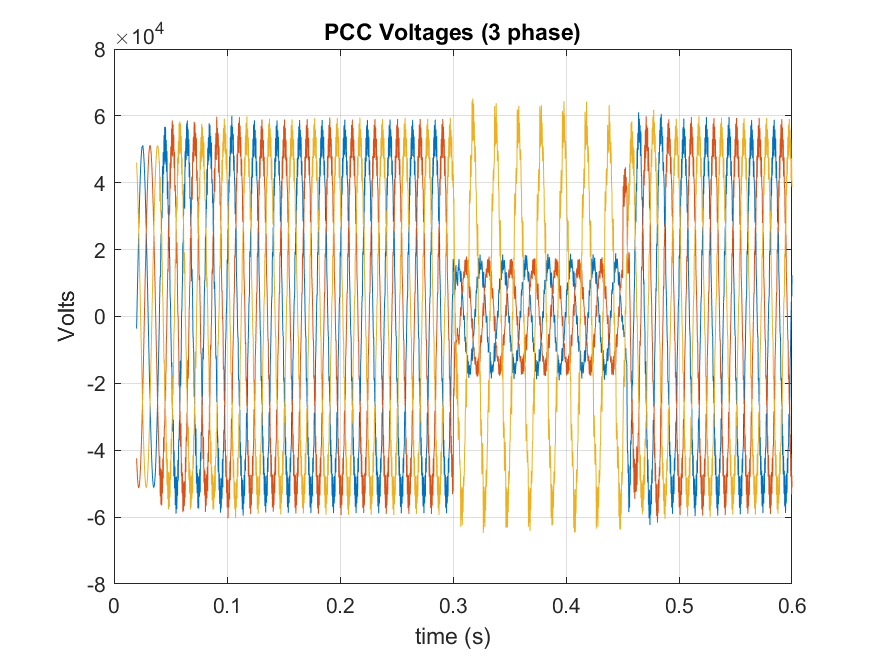}}
	\subfigure[]{\includegraphics[width=0.47\textwidth]{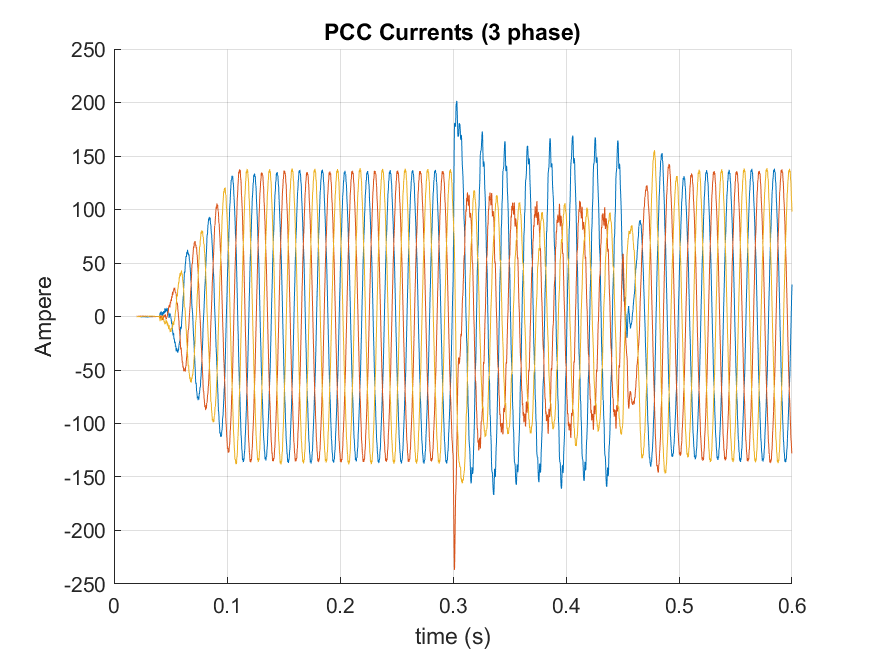}}\\
	\subfigure[]{\includegraphics[width=0.47\textwidth]{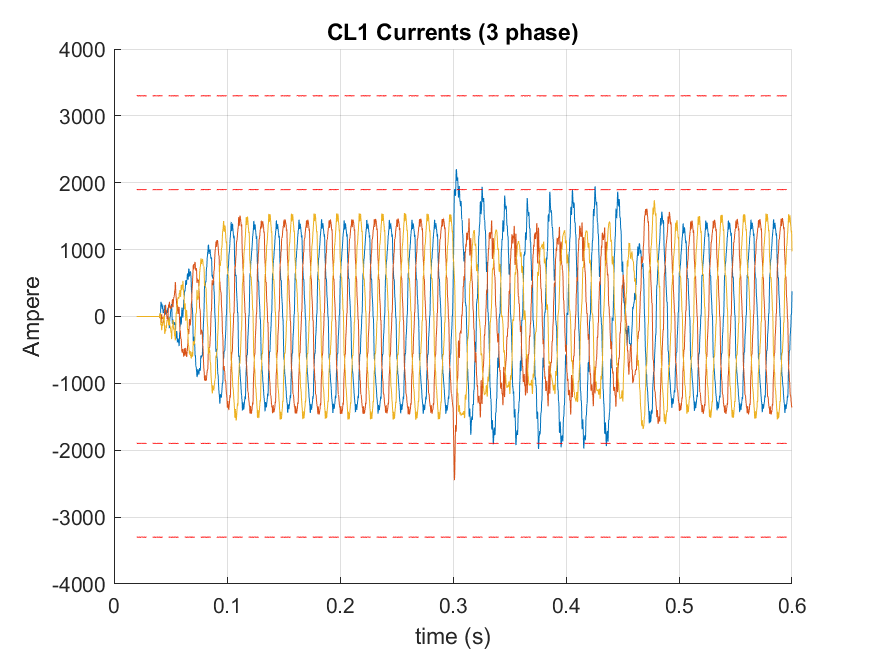}\label{fig::cl1}}
	\subfigure[]{\includegraphics[width=0.47\textwidth]{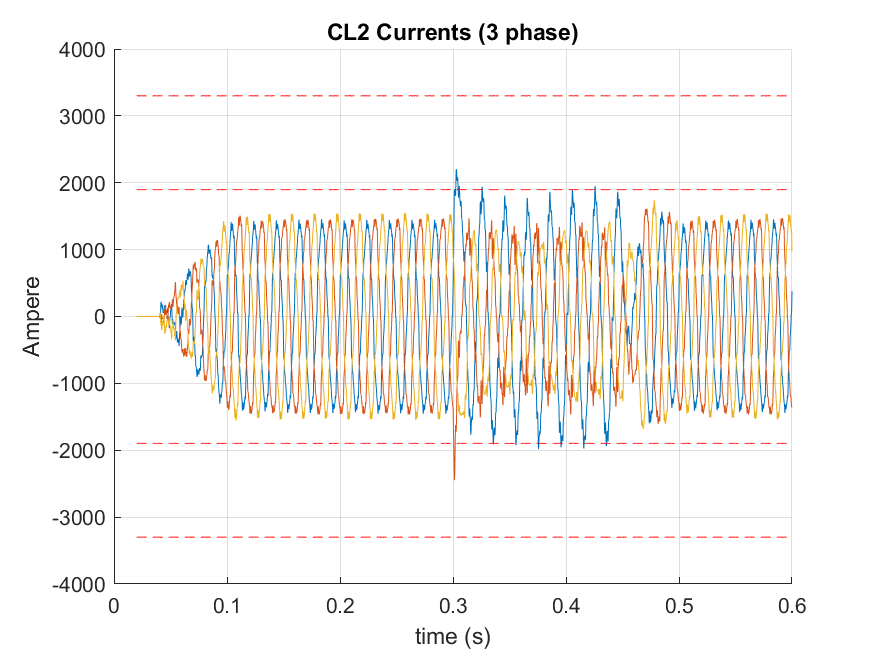}\label{fig::cl2}}
	\caption{Asymmetric 25\%, Container Fault}
\end{figure}

\begin{figure}[t]
	\centering
	\subfigure[]{\includegraphics[width=0.44\textwidth]{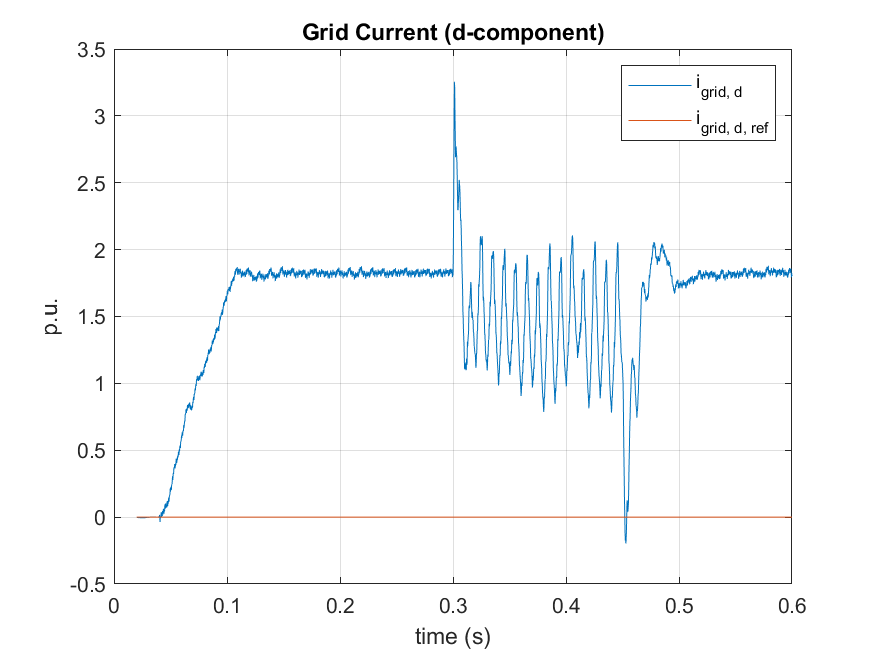}\label{fig::iPSd}}
	\subfigure[]{\includegraphics[width=0.44\textwidth]{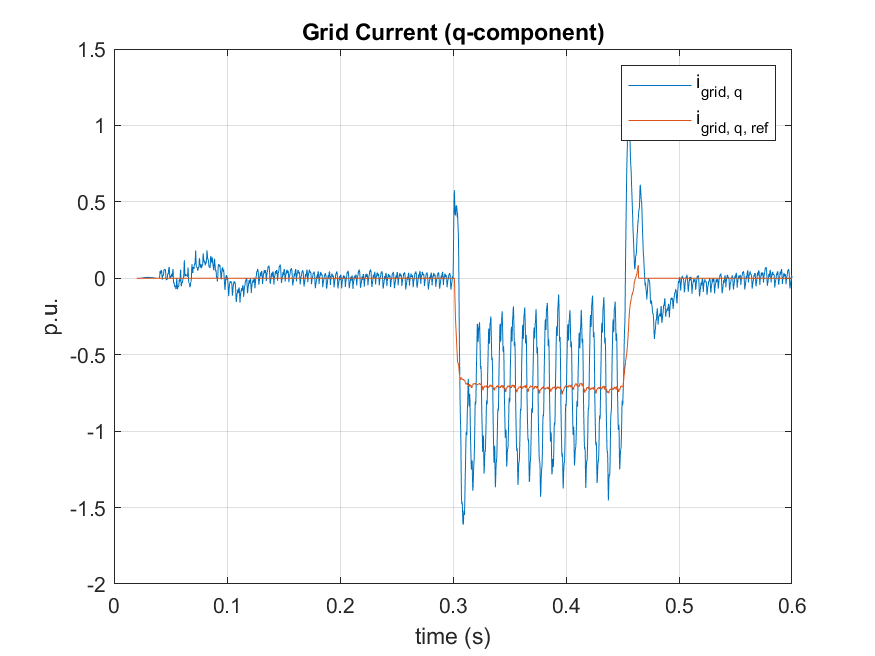}\label{fig::iPSq}}\\
	\subfigure[]{\includegraphics[width=0.44\textwidth]{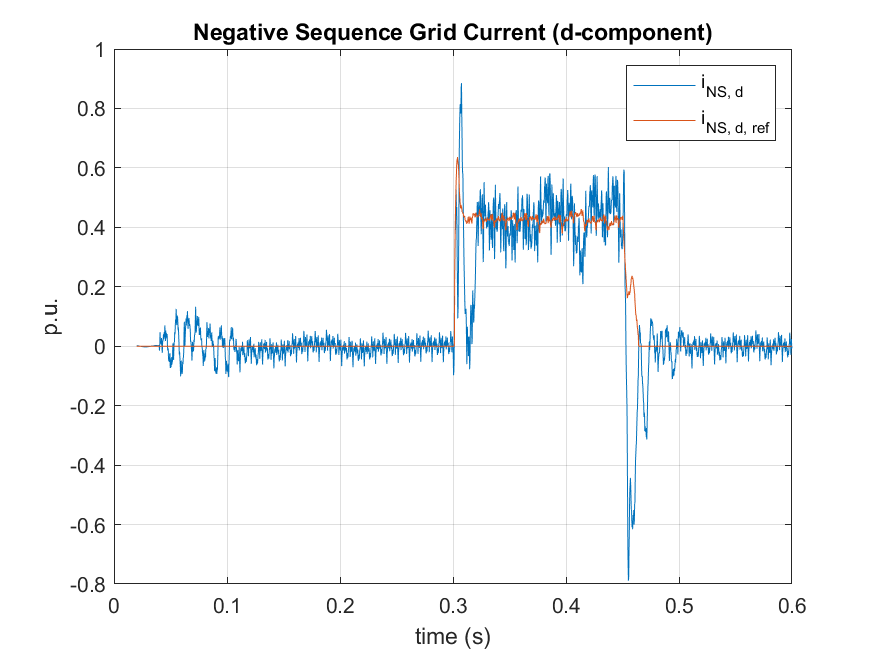}\label{fig::iNSd}}
	\subfigure[]{\includegraphics[width=0.44\textwidth]{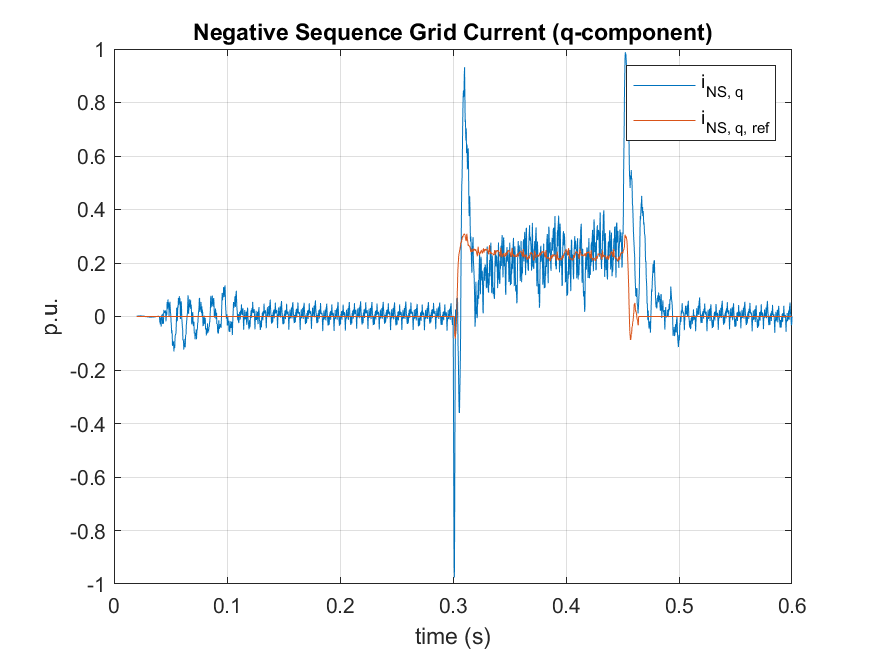}\label{fig::iNSq}}\\
	\subfigure[]{\includegraphics[width=0.44\textwidth]{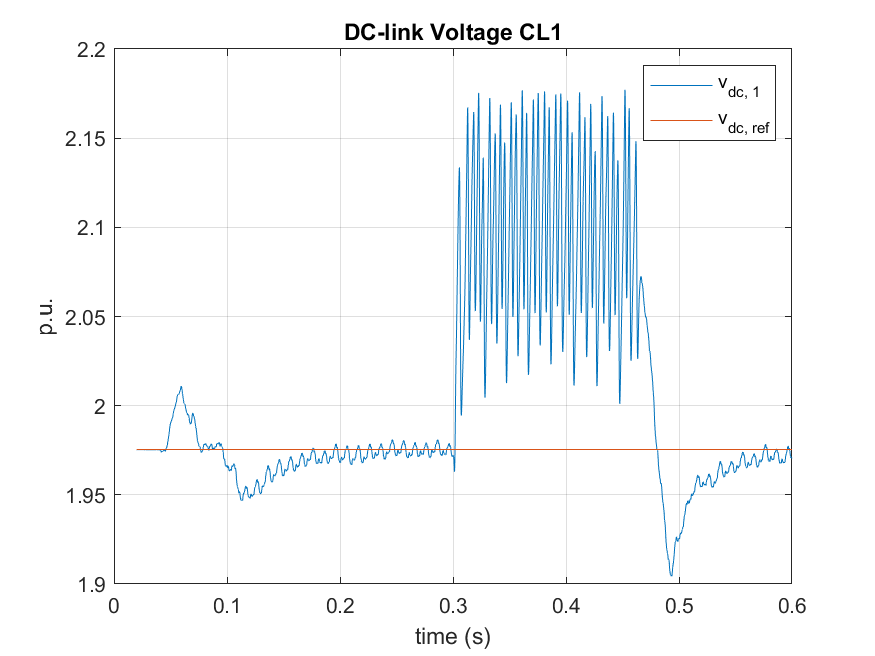}\label{fig::vdc1}}
	\subfigure[]{\includegraphics[width=0.44\textwidth]{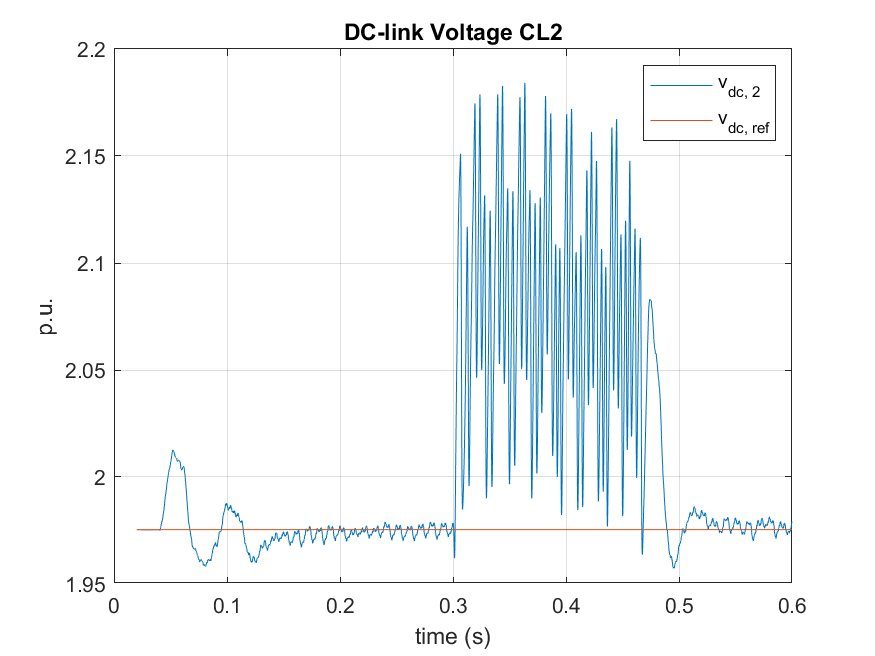}\label{fig::vdc2}}
	\caption{Asymmetric 25\%, Container Fault}
\end{figure}

\textbf{Asymmetric 25\%, Container Fault:}
We investigate how the system performs when an asymmetric fault in the container side brings the PCC voltage to 25\% of its original value in two out of the three phases at SCR 3. It can be observed in \cref{fig::cl1,fig::cl2} that the conversion line currents are kept within their operating limits and always inside the safe operating region. The first current limit corresponds to $\ol{i_{cl}}=\SI{1.2}{{\pu}}$ and the second one depicts $\SI{2}{{\pu}}$.
Moreover, the positive sequence $\qcom$-current reference in \cref{fig::iPSq} and the negative sequence $\dq$-current references in \cref{fig::iNSd,fig::iNSq} are closely tracked. The oscillations are observed in $\dq$-currents of \cref{fig::iPSd,fig::iPSq} because the negative sequence current components are not filtered out (they appear at twice the fundamental frequency). Observe that the dc-link voltage is balanced during the ramp-up, steady-state and fault-clearing phases, as shown in \cref{fig::vdc1,fig::vdc2}. During fault, voltage limiting unit is active, since it is not feasible to inject the generated power to the grid, while prioritizing reactive power reference. Note that all figures include the ripple components originating from the low switching frequency.

\begin{figure}[t]
	\centering
	\subfigure[]{\includegraphics[width=0.44\textwidth]{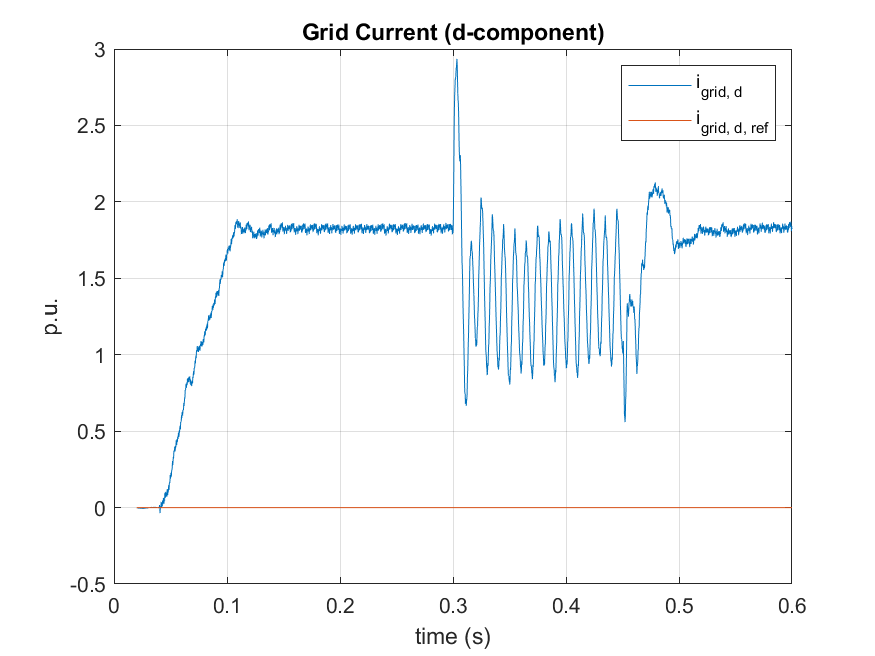}}
	\subfigure[]{\includegraphics[width=0.44\textwidth]{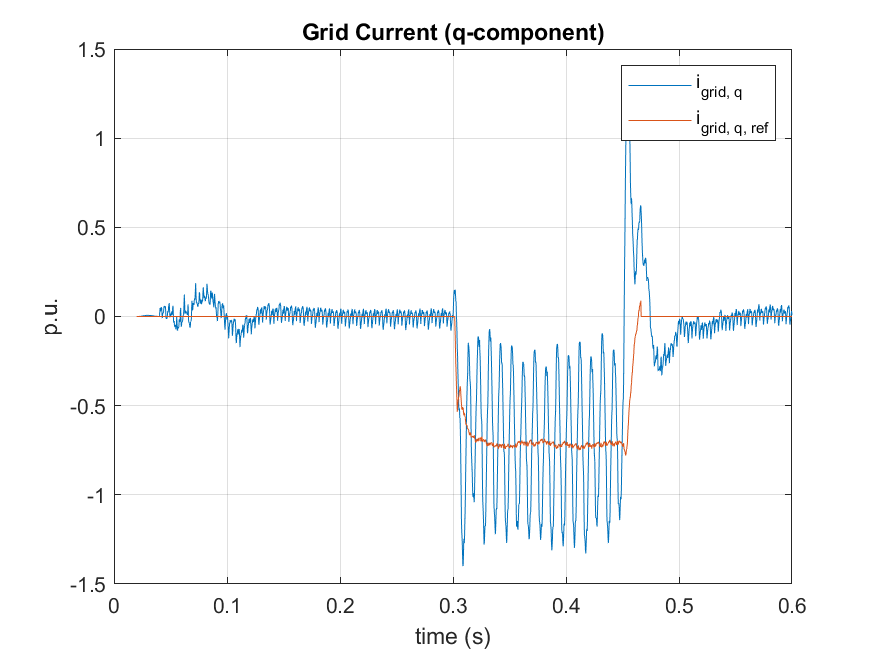}}\\
	\subfigure[]{\includegraphics[width=0.44\textwidth]{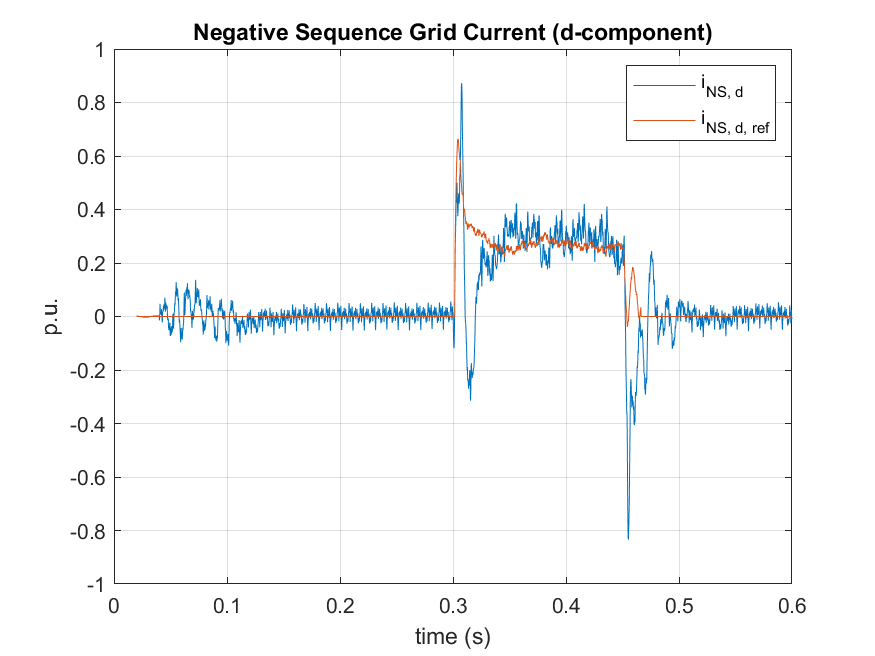}}
	\subfigure[]{\includegraphics[width=0.44\textwidth]{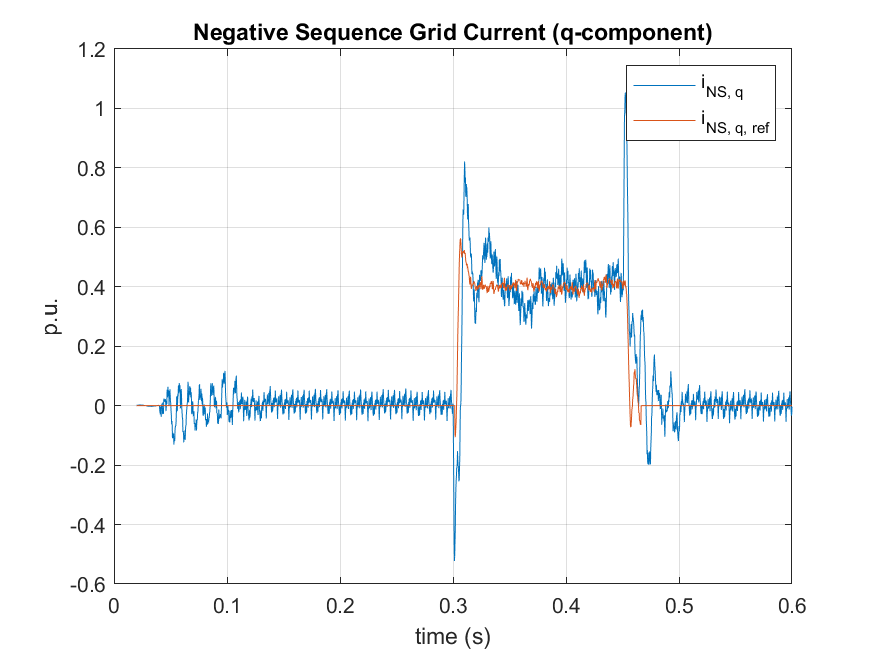}}\\
	\subfigure[]{\includegraphics[width=0.44\textwidth]{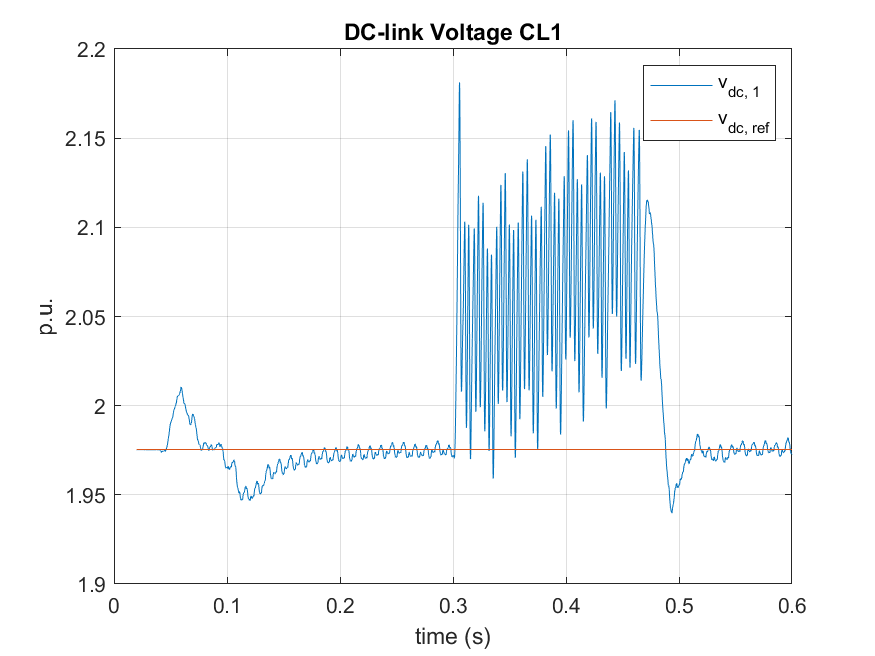}}
	\subfigure[]{\includegraphics[width=0.44\textwidth]{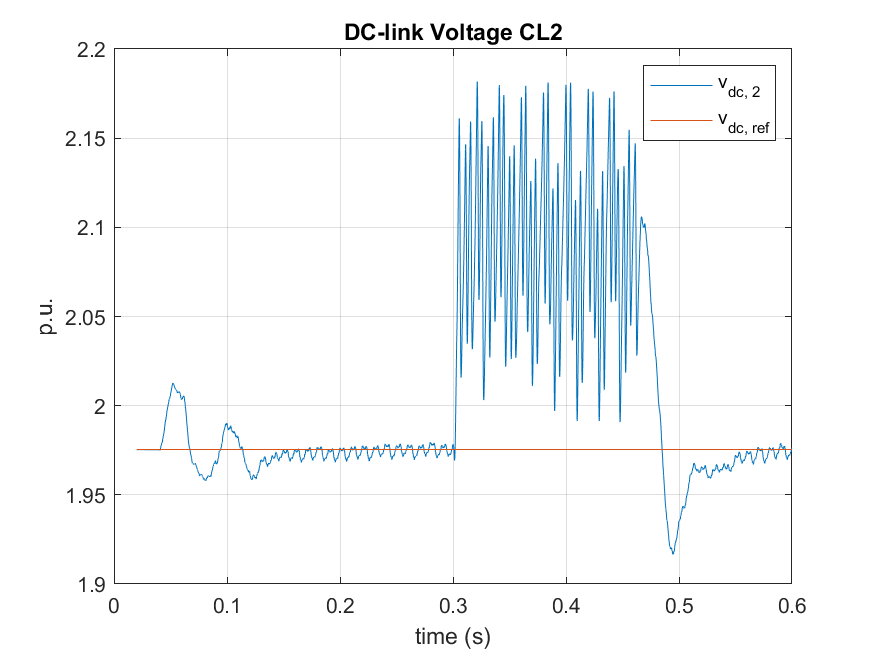}}
	\caption{Asymmetrical 1\%, Grid Source Fault}\label{fig:grid_source}
\end{figure}

\textbf{Asymmetric 1\%, Grid Source Fault:}
We investigate an asymmetrical fault in the source side that brings the (infinite-bus) grid voltage to 1\% of its nominal value in two out of the three phases. The conversion line currents are again kept between their operating limits and always inside the safe operating region. Moreover, the dc-link voltages are also balanced. Finally, the positive sequence $\qcom$-current and the negative sequence $\dq$-current references are closely tracked. For the tracking performance in this fault study, we refer to Figure~\ref{fig:grid_source}.


\section{Conclusions}

This work developed an MPC scheme for the grid-side control of wind power conversion systems. To obtain an optimization problem formulation that can be solved within a few hundreds of microseconds in a receding horizon fashion, we reduced the number of decision variables by restricting the control inputs (i.e., converter modulating signals) to a certain frequency content. Numerical case studies showed that such an MPC scheme can exhibit fast response in transients and fault tests, while keeping the converter always within its safe operating region. Future work could expand on this work with a larger set of case studies, and with an implementation on an embedded device. \looseness=-1

\section*{Acknowledgment}
The authors thank Tinus Dorfling, Ioannis Tsoumas and ABB System Drives colleagues for their inputs for this work.\looseness=-1

\bibliographystyle{IEEEtran}
\bibliography{report}

\begin{thebibliography}{10}
\providecommand{\url}[1]{#1}
\csname url@samestyle\endcsname
\providecommand{\newblock}{\relax}
\providecommand{\bibinfo}[2]{#2}
\providecommand{\BIBentrySTDinterwordspacing}{\spaceskip=0pt\relax}
\providecommand{\BIBentryALTinterwordstretchfactor}{4}
\providecommand{\BIBentryALTinterwordspacing}{\spaceskip=\fontdimen2\font plus
\BIBentryALTinterwordstretchfactor\fontdimen3\font minus
  \fontdimen4\font\relax}
\providecommand{\BIBforeignlanguage}[2]{{%
\expandafter\ifx\csname l@#1\endcsname\relax
\typeout{** WARNING: IEEEtran.bst: No hyphenation pattern has been}%
\typeout{** loaded for the language `#1'. Using the pattern for}%
\typeout{** the default language instead.}%
\else
\language=\csname l@#1\endcsname
\fi
#2}}
\providecommand{\BIBdecl}{\relax}
\BIBdecl

\bibitem{tsoumas2023elimination}
I.~Tsoumas and L.~Harnefors, ``Elimination of back-emf induced dc-link voltage
  ripple in wecs with permanent magnet synchronous generators,'' in
  \emph{EPE'23 ECCE Europe}.\hskip 1em plus 0.5em minus 0.4em\relax IEEE, 2023,
  pp. 1--7.

\bibitem{karamanakos2020model}
P.~Karamanakos, E.~Liegmann, T.~Geyer, and R.~Kennel, ``Model predictive
  control of power electronic systems: Methods, results, and challenges,''
  \emph{IEEE Open J. of In. Appl.}, vol.~1, pp. 95--114, 2020.

\bibitem{harbi2023model}
I.~Harbi, J.~Rodriguez, E.~Liegmann, H.~Makhamreh, M.~L. Heldwein, M.~Novak,
  M.~Rossi, M.~Abdelrahem, M.~Trabelsi, M.~Ahmed \emph{et~al.}, ``Model
  predictive control of multilevel inverters: Challenges, recent advances, and
  trends,'' \emph{IEEE Trans. on Power Elec.}, 2023.

\bibitem{rodriguez2021latest}
J.~Rodriguez, C.~Garcia, A.~Mora, F.~Flores-Bahamonde, P.~Acuna, M.~Novak,
  Y.~Zhang, L.~Tarisciotti, S.~A. Davari, Z.~Zhang \emph{et~al.}, ``Latest
  advances of model predictive control in electrical drives—part i: Basic
  concepts and advanced strategies,'' \emph{IEEE Trans. on Power Elec.},
  vol.~37, no.~4, pp. 3927--3942, 2021.

\bibitem{rodriguez2021latest2}
J.~Rodriguez, C.~Garcia, A.~Mora, S.~A. Davari, J.~Rodas, D.~F. Valencia,
  M.~Elmorshedy, F.~Wang, K.~Zuo, L.~Tarisciotti \emph{et~al.}, ``Latest
  advances of model predictive control in electrical drives—part ii: App. and
  benchmarking with classical control methods,'' \emph{IEEE Trans. on Power
  Elec.}, vol.~37, no.~5, pp. 5047--5061, 2021.

\bibitem{zafra2023long}
E.~Zafra, S.~Vazquez, T.~Geyer, R.~P. Aguilera, and L.~G. Franquelo, ``Long
  prediction horizon fcs-mpc for power converters and drives,'' \emph{IEEE Open
  J. of the In. Elec. Soc.}, 2023.

\bibitem{dorfling2022generalized}
T.~Dorfling, H.~du~Toit~Mouton, and T.~Geyer, ``Generalized model predictive
  pulse pattern control based on small-signal modeling—part 1: Algorithm,''
  \emph{IEEE Trans. on Power Elec.}, vol.~37, no.~9, pp. 10\,476--10\,487,
  2022.

\bibitem{dorfling2022generalized2}
------, ``Generalized model predictive pulse pattern control based on
  small-signal modeling—part 2: Implementation and analysis,'' \emph{IEEE
  Trans. on Power Elec.}, vol.~37, no.~9, pp. 10\,488--10\,498, 2022.

\bibitem{begh2022gradient}
M.~A.~W. Begh, P.~Karamanakos, and T.~Geyer, ``Gradient-based predictive pulse
  pattern control of medium-voltage drives—part i: Control, concept, and
  analysis,'' \emph{IEEE Trans. on Power Elec.}, vol.~37, no.~12, pp.
  14\,222--14\,236, 2022.

\bibitem{begh2022gradient2}
M.~A.~W. Begh, P.~Karamanakos, T.~Geyer, and Q.~Yang, ``Gradient-based
  predictive pulse pattern control of medium-voltage drives—part ii:
  Performance assessment,'' \emph{IEEE Trans. on Power Elec.}, vol.~37, no.~12,
  pp. 14\,237--14\,251, 2022.

\bibitem{bolognani2008design}
S.~Bolognani, S.~Bolognani, L.~Peretti, and M.~Zigliotto, ``Design and
  implementation of model predictive control for electrical motor drives,''
  \emph{IEEE Trans. on In. Elec.}, vol.~56, no.~6, pp. 1925--1936, 2008.

\bibitem{mariethoz2008explicit}
S.~Mari{\'e}thoz and M.~Morari, ``Explicit model-predictive control of a {PWM}
  inverter with an {LCL} filter,'' \emph{IEEE Trans. on In. Elec.}, vol.~56,
  no.~2, pp. 389--399, 2008.

\bibitem{darivianakis2014model}
G.~Darivianakis, T.~Geyer, and W.~van~der Merwe, ``Model predictive current
  control of modular multilevel converters,'' in \emph{ECCE}.\hskip 1em plus
  0.5em minus 0.4em\relax IEEE, 2014, pp. 5016--5023.

\bibitem{rossi2022indirect}
M.~Rossi, P.~Karamanakos, and F.~Castelli-Dezza, ``An indirect model predictive
  control method for grid-connected three-level neutral point clamped
  converters with {LCL} filters,'' \emph{IEEE Trans. on In. App.}, vol.~58,
  no.~3, pp. 3750--3768, 2022.

\bibitem{keusch2023long}
R.~Keusch, H.-A. Loeliger, and T.~Geyer, ``Long-horizon direct model predictive
  control for power converters with state constraints,'' \emph{IEEE Trans. on
  Contr. Syst. Tech.}, 2023.

\bibitem{karamanakos2014direct}
P.~Karamanakos, T.~Geyer, N.~Oikonomou, F.~D. Kieferndorf, and S.~Manias,
  ``Direct model predictive control: A review of strategies that achieve long
  prediction intervals for power electronics,'' \emph{IEEE In. Elec. Mag.},
  vol.~8, no.~1, pp. 32--43, 2014.

\bibitem{karttunen2013decoupled}
J.~Karttunen, S.~Kallio, P.~Peltoniemi, P.~Silventoinen, and O.~Pyrh{\"o}nen,
  ``Decoupled vector control scheme for dual three-phase permanent magnet
  synchronous machines,'' \emph{IEEE Trans. on In. Elec.}, vol.~61, no.~5, pp.
  2185--2196, 2013.

\bibitem{teodorescu_grid_2011}
R.~Teodorescu, M.~Liserre, and P.~Rodr{\'i}guez, ``Grid converters for
  photovoltaic and wind power systems,'' 2011.

\bibitem{kaura1997operation}
V.~Kaura and V.~Blasko, ``Operation of a phase locked loop system under
  distorted utility conditions,'' \emph{IEEE Trans. on In. Appl.}, vol.~33,
  no.~1, pp. 58--63, 1997.

\bibitem{ho2013method}
N.-M. Ho, G.~Escobar, and S.~Pettersson, ``Method and arrangement for detecting
  frequency and fundamental wave component of three-phase signal,'' 2013, uS
  Patent App. 13/666,649.

\bibitem{lofberg2004yalmip}
J.~Lofberg, ``Yalmip: A toolbox for modeling and optimization in matlab,'' in
  \emph{IEEE ICRA}, 2004, pp. 284--289.

\bibitem{banjac2017embedded}
G.~Banjac, B.~Stellato, N.~Moehle, P.~Goulart, A.~Bemporad, and S.~Boyd,
  ``Embedded code generation using the osqp solver,'' in \emph{CDC}.\hskip 1em
  plus 0.5em minus 0.4em\relax IEEE, 2017, pp. 1906--1911.

\bibitem{stellato2020osqp}
B.~Stellato, G.~Banjac, P.~Goulart, A.~Bemporad, and S.~Boyd, ``Osqp: An
  operator splitting solver for quadratic programs,'' \emph{Math. Prog. Comp.},
  vol.~12, no.~4, pp. 637--672, 2020.

\end{thebibliography}

\end{document}